\documentclass{vgtc}                          

\graphicspath{{figures/}{pictures/}{images/}{./}} 

\usepackage{times}                     

\usepackage{tabu}                      
\usepackage{booktabs}                  
\usepackage{lipsum}                    
\usepackage{mwe}                       

\usepackage{mathptmx}                  

\usepackage{subcaption}

\onlineid{0}

\vgtccategory{Research}

\vgtcinsertpkg

\hypersetup{
    pdftitle={Toward Understanding Similarity of Visualization Techniques},
    pdfauthor={Abdulhaq Adetunji Salako and Christian Tominski},
    pdfsubject={Taking first steps toward understanding similarity of visualization techniques},
    pdfkeywords={similarity, visualization, data facets},
}

\title{Toward Understanding Similarity of Visualization Techniques}

\author{Abdulhaq Adetunji Salako\thanks{e-mail: abdulhaq.salako@uni-rostock.de} %
\and Christian Tominski\thanks{e-mail:christian.tominski@uni-rostock.de}}
\affiliation{\scriptsize University of Rostock \\ Institute of Visual and Analytic Computing \\Rostock, Germany}

\teaser{
  \centering
  \includegraphics[width=\textwidth]{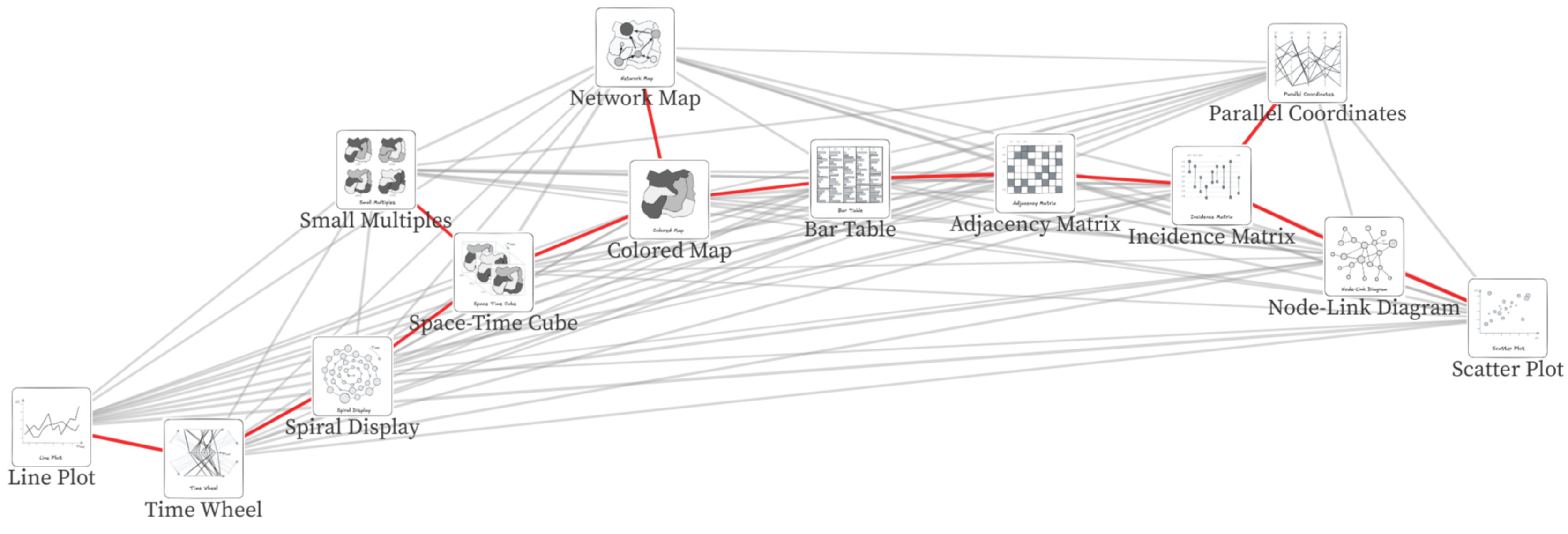}
  \caption{Minimum spanning tree illustrating the strongest model-driven similarity between visualization techniques.}
  \label{fig:mst_model}
}

\abstract{
    The literature describes many visualization techniques for different types of data, tasks, and application contexts, and new techniques are proposed on a regular basis. Visualization surveys try to capture the immense space of techniques and structure it with meaningful categorizations. Yet, it remains difficult to understand the similarity of visualization techniques in general. We approach this open research question from two angles. First, we follow a model-driven approach that is based on defining the signature of visualization techniques and interpreting the similarity of signatures as the similarity of their associated techniques. Second, following an expert-driven approach, we asked visualization experts in a small online study for their ad-hoc intuitive assessment of the similarity of pairs of visualization techniques. From both approaches, we gain insight into the similarity of a set of 13 basic and advanced visualizations for different types of data. While our results are so far preliminary and academic, they are first steps toward better understanding the similarity of visualization techniques.
} 


\begin{document}



\maketitle


\section{Introduction}
\label{sec:intro}

The visualization community has created a large number of visualization techniques. Visualization surveys document the wide variety of existing techniques. For example, a survey on tree visualization \cite{Schulz11Tree} lists more than 300 works, a survey on visualization of time-oriented data \cite{Aigner23TimeViz2nd} covers more that 160 techniques, and about the same number of works is collected in a survey of dynamic graph visualization \cite{Beck14DynGraphVis}. 

This variety of techniques is both a curse and a blessing. A curse, because it can be difficult to find a suitable visualization technique (or determine the lack thereof) in the vast space of options. A blessing, because there are solutions for many different types of data, analysis tasks, and application scenarios.

To better understand the space of visualization techniques, it would be good to understand the similarity of visualization techniques. Yet the notion of similarity is not easy to capture and might involve various aspects. Recent previous work has explored an image-based typology of visualizations, emphasizing the role of visual appearance and perceptual properties of visual representations \cite{Chen24Typology}. Going beyond plain appearance, existing visualization surveys group previous works based on structured categorization schemes. The tree visualization survey \cite{Schulz11Tree}, for example, distinguishes explicit, implicit, and hybrid representations, as one among three category dimensions. The TimeViz Browser \cite{Tominski23TimeVizBrowser} groups, for example, 2D and 3D as well as static and dynamic visualization techniques. 
While such categorizations help us structure the space of existing works, they cannot fully capture the similarity of visualization techniques. On the one hand, a categorization group can still contain techniques that are quite heterogeneous in terms of their visual appearance. On the other hand, surveys may list distinct techniques that appear to be so similar that they should be considered variants of the same technique, rather than unique techniques. For example, the TimeViz Browser \cite{Tominski23TimeVizBrowser} lists spiral graph, spiral display, and enhanced interactive spiral as visualization techniques, but they are essentially all variants of the same technique.

If we had a good understanding of the similarity of visualization techniques, we could better identify such cases and probably condense the big zoo of techniques to a more compact space. This would make it easier for scientists to identify research gaps, and for practitioners to find the techniques that fit their data analysis problems. Knowing about the similarity of techniques can also have implications in other contexts. For example, similarity-based interfaces could be created to enable the visual exploration of the space of visualization techniques \cite{Schulz15Preset}. Also the teaching of visualization techniques could benefit from introducing (unique) techniques in order of their similarity. Students could learn by analogy and easily transfer knowledge about an already learned technique to the next (similar) technique to be learned \cite{Ruchikachorn15Learning}. Similarly, multi-view visualization systems \cite{Chen21MultiView} could arrange their views such that similar techniques are shown next to each other to make it easier for users to transfer insight from one view to another. This is particularly relevant when visualizing multi-faceted data, where usually each data facet needs to be represented with a dedicated visualization technique \cite{Hadlak15multifaceted}.

This work presents our preliminary and ongoing research toward understanding similarity of visualization techniques. We explore visualization similarity from two perspectives. First, a \emph{model-driven perspective} focuses on the \emph{structural and semantic properties} of a visualization. These properties are encoded in a string signature, and the similarity of visualization techniques is derived based on the similarity of their signatures. Second, an \emph{expert-driven perspective} captures the \emph{ad-hoc intuitive human understanding} of similarity of three visualization experts. From both studies, we can learn about the similarities of a curated set of 13 visualization techniques covering four different data facets. We understand our research as stepping stones toward more substantial future research on understanding the similarity of visualization techniques, which we think can have substantial theoretical and practical value.
\section{Toward Similarity of Visualization Techniques}
\label{sec3}

In this section, we describe the set of visualization techniques as well as our model-driven and expert-driven approaches in detail. 


{
\renewcommand{\arraystretch}{1.2}%
\begin{table}[t]
    \centering
    \caption{Visualization techniques and their signatures.}
    \label{table:sequence_legend}
    \scriptsize%
	\centering%
    \begin{tabu}{ll}
        \toprule
        \textbf{Vis. Technique} & \textbf{Signature} \\
        \midrule
        Bar Table \textbf{BT}            & $D_AM_AC_PC_LC_CR_AO_LL_S$ \\
        Scatter Plot \textbf{SP}         & $D_AM_PC_PC_AC_CR_SO_LL_D$ \\
        Parallel Coordinates \textbf{PC} & $D_AM_LC_PC_CR_AO_PL_D$ \\
        Line Plot \textbf{LP}            & $D_TD_AM_PM_LC_PC_CR_OO_LL_D$ \\
        Spiral Display \textbf{SD}       & $D_TD_AM_AC_PC_CR_OO_RL_S$ \\
        Time Wheel \textbf{TW}           & $D_TD_AM_LC_PC_CR_OO_RL_D$ \\
        Colored Map \textbf{CM}          & $D_SD_AM_AC_PC_CC_SR_SO_LL_S$ \\
        Small Multiples \textbf{SM}      & $D_TD_SM_PM_LM_AC_PR_AO_LO_PO_RL_S$ \\
        Space-Time Cube \textbf{STC}     & $D_TD_SM_AC_PC_CC_SR_OO_LL_S$ \\
        Network Map \textbf{NM}          & $D_SD_RM_PM_LM_AC_PC_AC_CC_SR_SR_OO_LL_S$ \\
        Node-Link Diagram \textbf{NLD}   & $D_RD_AM_PM_LC_PC_AC_CR_SO_LL_D$ \\
        Adjacency Matrix \textbf{AM}     & $D_RD_AM_AC_PC_CR_AO_LL_S$ \\
        Incidence Matrix \textbf{IM}     & $D_RD_AM_LC_PC_AC_CR_AO_LO_PL_D$ \\
        \bottomrule
    \end{tabu}
\end{table}
}

\subsection{Technique Collection}

Given the vast variety of visualization techniques, a first necessary step is to define a collection of techniques that is both representative and of manageable size. Our goal was to include techniques that cover multiple data facets, classic and contemporary, as well as basic and advanced techniques of different visual appearances. The number of techniques should be reasonably small to make the planned expert study feasible. For the different data facets, we considered visualization techniques that were designed for representing time (T), space (S), multivariate attributes (A), and structural relationships (R) as defined in \cite{Tominski20IVDA}. 

\begin{figure}[t]
 \centering
 \includegraphics[width=\columnwidth]{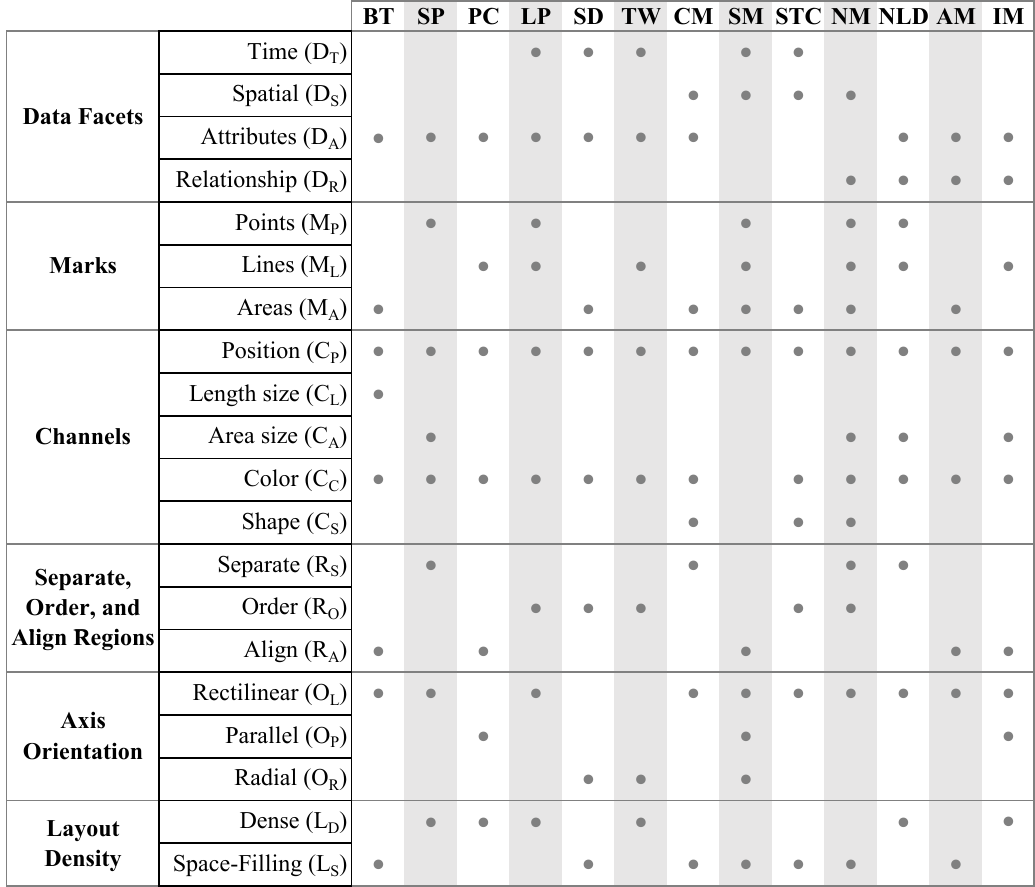}
 \caption{Categorization of visualization techniques.}
 \label{fig:technique_corpus}
\end{figure}

Based on the T, S, A, and R data facets and existing surveys, we extracted visualization techniques that support ideally two, but at least one of these facets. Through a refinement process, we curated a final corpus of 13 distinct visualization techniques as listed in \cref{table:sequence_legend} and \cref{fig:technique_corpus}, with at least three techniques for each data facet, and also including the desired classic and contemporary as well as basic and advanced techniques.





\subsection{Model-Driven Approach}

The goal of the model-driven approach is to systematically assess the similarity of visualization techniques based on some abstracted or derived model. In other words, we would consider two visualization techniques to be similar if their associated models are similar.

One can consider several alternatives to abstractly model a visualization technique. One option could be grammar-based specifications, such as Vega-Lite specifications. A second option could be to consider the scene graph underlying a visual representation, for example, an SVG representation. The similarity of visualization techniques could then be determined based on the structural similarity of Vega-Lite specifications or SVG scene graphs. While these approaches could work well for basic charts such as line plots or scatter plots, they fall short for more complex techniques that are not covered by a grammar or are overly complicated to generate. Moreover, a substantial amount of work would be required to manually (re)create each visualization before models can be extracted.

\paragraph{Technique Signatures}

As a more scalable alternative, we adopted a simplified approach inspired by biological sequencing and the categorizations provided in existing surveys. The goal of the sequencing is to generate technique \emph{signatures} that should ideally capture the key structural and semantic properties of a visualization technique. Again, there are many different properties that could be taken into account. Based on Munzner's book \cite{Munzner14VAD}, we considered what data facets are represented, what marks and channels are used, or how the marks are oriented and aligned.

With the properties listed in \cref{fig:technique_corpus}, we encoded each technique as a sequence of categorical tokens representing a technique's signature. Each signature begins with one or two data facet tokens, representing the primary and, if applicable, the secondary data facet that a technique was designed to visualize. For example, a Node-Link Diagram (\textbf{NLD}) primarily visualizes the relational structure ($D_R$) of a network and can also represent additional multivariate attributes ($D_A$) of nodes and edges. The data facet tokens are followed by tokens describing the visual encoding (i.e., marks and channels). For example, Parallel Coordinates (\textbf{PC}) use line marks ($M_L$) and position ($C_P$) and optionally color ($C_C$) as visual channels. Finally, signatures contain tokens for extras such as axis orientation or layout density as defined by Munzner \cite{Munzner14VAD}. For example, a Spiral Display (\textbf{SD}) has a radial axis orientation ($O_R$), and for an Adjacency Matrix (\textbf{AM}), the layout density is space-filling ($L_S$).



For a complete example, consider the signature for the Line Plot (\textbf{LP}) technique $D_T D_A M_P M_L C_P C_C R_O O_L L_D$, capturing its key design features. It begins with $D_T D_A$, indicating a focus on \emph{time} and \emph{attributes} as data facets. The $M_P M_L$ tokens show that \textbf{LP} uses \emph{points} and \emph{lines} as visual marks. The channel tokens $C_P C_C$ denote the use of \emph{position} and \emph{color} for the visual data encoding. $R_O$ signifies that the data are \emph{ordered}, while $O_L$ and $L_D$ describe a \emph{rectilinear} and \emph{dense} layout. Such signatures form the basis for computing pairwise similarity using sequence alignment and distance metrics.

It should be noted that our signature approach is of course a compromise that aims to balance the modeling effort and the expressiveness of the obtained signatures. Moreover, certain parts of the signature are not necessarily unique per technique. For example, Line Plots could also vary the line width or Scatter Plots could use various shapes as additional visual channels to visualize another multivariate data attribute. Being aware of these limitations, we generated the 13 signatures as shown in \cref{table:sequence_legend}.

\paragraph{Signature Similarity} 

Assuming that signatures represent visualization techniques reasonably well, visualization similarity can be approximated based on the signatures. This corresponds to quantifying the pairwise difference between two signatures. To this end, we can employ string-based distance metrics, which allow us to measure structural dissimilarity. Among various available metrics, we selected the Jaro-Winkler similarity metric \cite{Winkler90String}, originally developed for identifying duplicate records in datasets based on name similarity. This metric is particularly well-suited for comparing categorical sequences of varying lengths, making it appropriate for our technique signatures. As a result, we obtain a normalized similarity score for each signature pair, supporting the overall analysis of similarity across our collected visualization techniques. The results obtained from our signature-based similarity approach are presented in \cref{subsec:sign-results}.


\subsection{Expert-Driven Approach}

In addition to the model-driven notion of similarity, we are also interested in an \emph{intuitive} understanding of similarity. Such an intuitive understanding could help us better understand the elusive concept of visualization similarity and its diverse influencing factors. To satisfy this interest, we asked experts about their \emph{ad-hoc} impression of the similarity of visualization techniques. To avoid confounding intuition with implementation details, we created sketchy-style drawings of each visualization technique in our corpus using the \href{https://excalidraw.com/}{Excalidraw} platform. These drawings were presented in pairs to visualization experts.

A small total of three experts participated in our online study. Two experts have over 20 years of experience in visualization research and teaching, while the third has approximately six years of academic and practical experience in the field. 

During the study, the experts evaluated pairs of techniques from our collection, rating their perceived similarity on a 5-point Likert scale in response to the question: ``How similar do you find these pairs of visualization techniques?'' All possible pairwise combinations were presented one after the other, resulting in \[{13 \choose 2} = 78\] unique pairs. For each pair of techniques, a response was recorded on a linear scale ranging from 1 (highly dissimilar) to 5 (highly similar). Overall, we obtained 234 similarity ratings. To complement these quantitative ratings, we introduced open-ended follow-up questions at random intervals, prompting participants to explain: ``Which aspects of this pair of techniques do you find most dis/similar?'' The corresponding qualitative responses are expected to offer deeper insight into the intuition and reasoning behind the experts' similarity judgments. The results of the expert study will be given in \cref{subsec:exp-results}.

It should be noted that 78 is already a large number of pairs to be rated. This is the reason why we initially limited the size of our corpus of techniques to only 13 and the number of experts in our preliminary study to only 3. Apparently, it would be desirable to study a larger corpus, say of 20 techniques, and get ratings by at least a dozen of experts. They then had to rate 190 pairs, for an overall of 2280 ratings. This would be a substantial amount of work, but it could also lead to more reliable results.

\section{Preliminary Findings}
\label{sec:findings}


\begin{figure*}[t]
    \includegraphics[width=\textwidth]{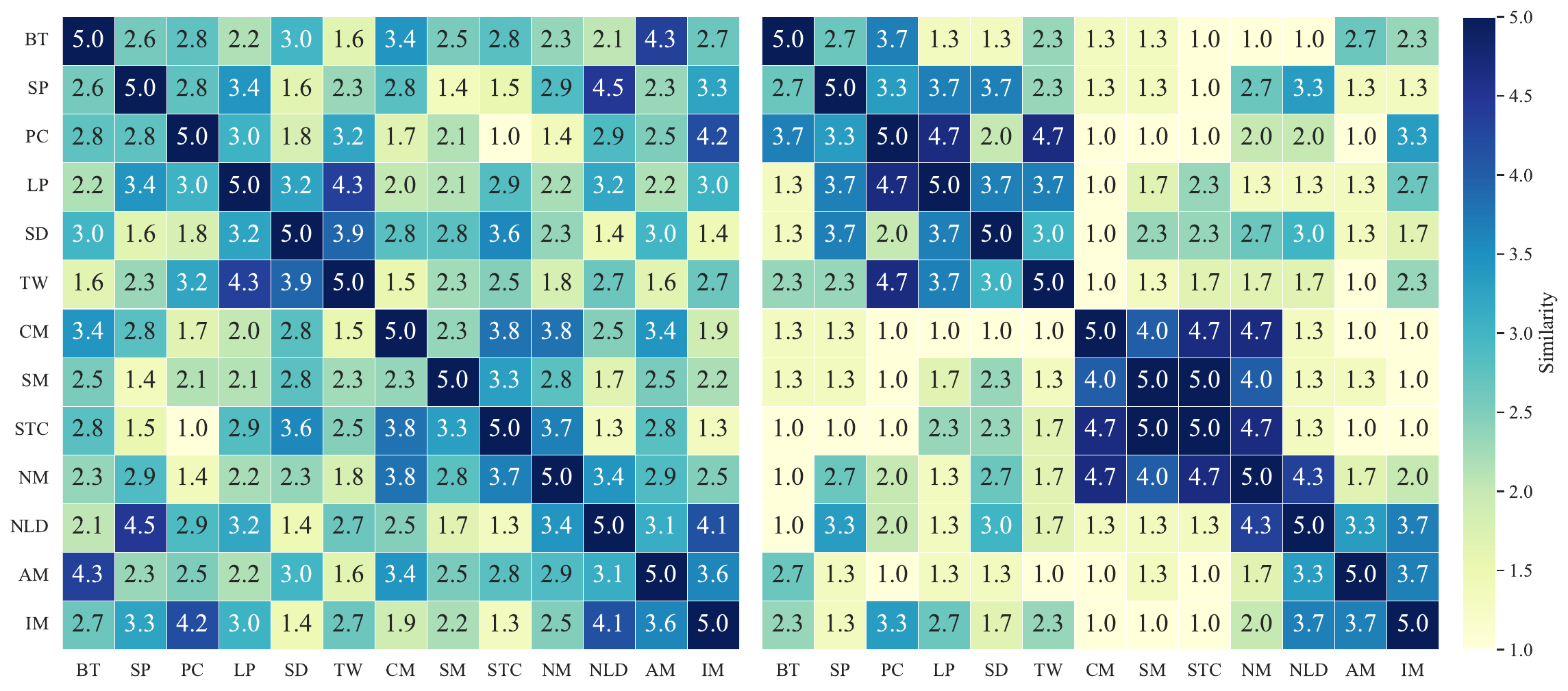}
	\caption{Model-driven (left) and expert-driven (right) similarity.}
	\label{fig:sim-model-expert}
\end{figure*}

\begin{figure}[t]
    \includegraphics[width=\columnwidth]{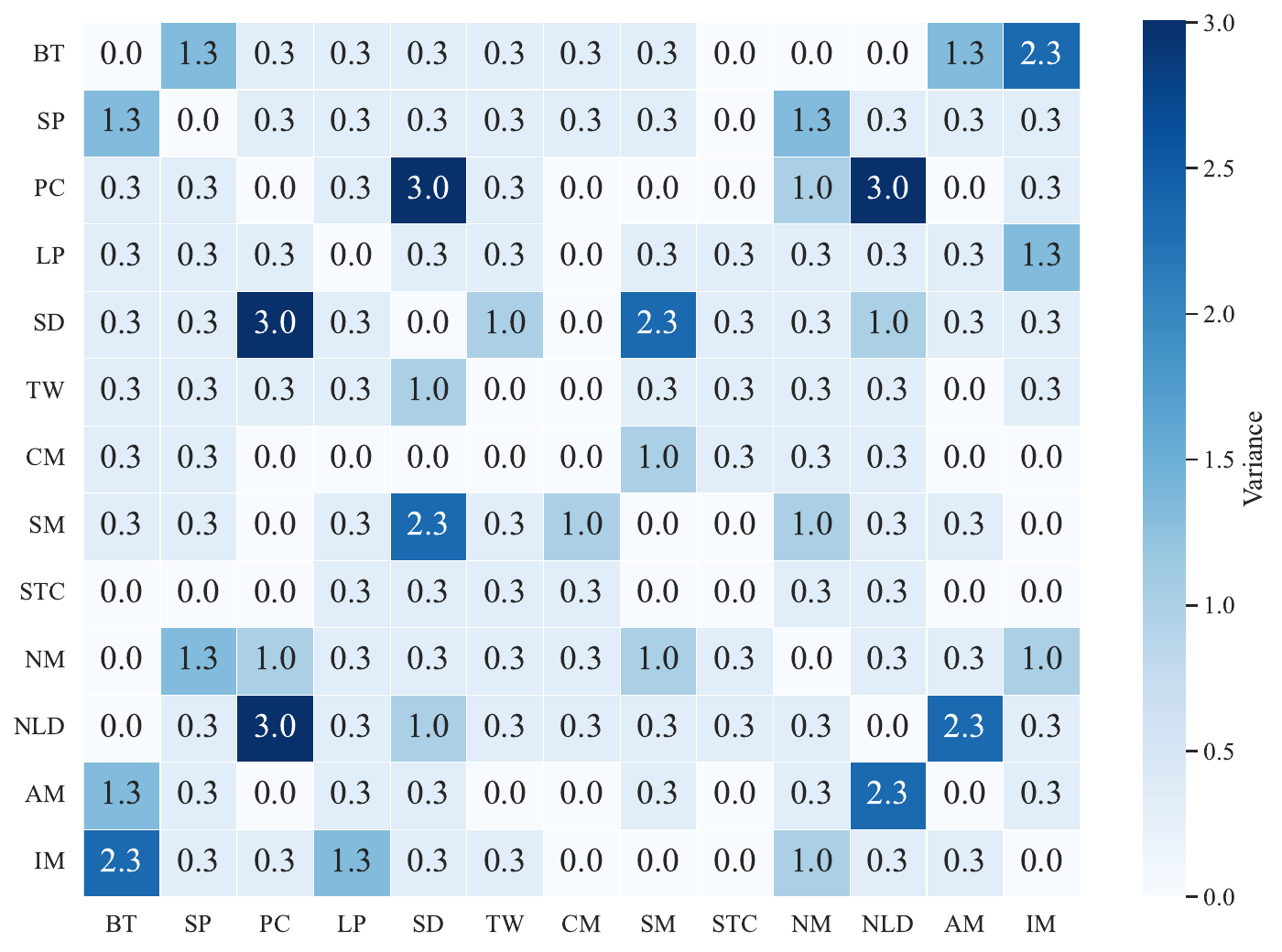}
	\caption{Variance among expert similarity ratings.}
	\label{fig:var-expert}
\end{figure}

From our model-driven and expert-driven approaches, we could obtain the following interesting results about the similarity of visualization techniques.

\subsection{Similarity based on Signatures}
\label{subsec:sign-results}


A heatmap visualization of the pairwise similarity scores generated from the model-based signature analysis is shown in \cref{fig:sim-model-expert} (left). Each cell represents the similarity score between two visualization techniques based on the Jaro-Winkler metric. Light yellow indicates low similarity and dark blue stands for high similarity. The original metric scores in the range of $[0.0-1.0]$ were scaled to $[1.0-5.0]$ for better comparability with the expert-driven results.

We can observe that techniques sharing common data facets generally exhibit higher similarity scores. For example, Node-Link Diagram (NLD), Adjacency Matrix (AM), and Incidence Matrix (IM) are rather similar from our model-driven perspective. However, these similarities often coexist with, or are even rivaled by, similarities due to common visual structures such as mark types, visual encoding, or layout. For instance, Adjacency Matrix (AM) also has strong similarity with Bar Table (BT) and Colored Map (CM). This similarity emerges from shared visual features like space-filling layout and area-based marks, despite differing data facets (relationships R vs. attributes A vs. space S) being visualized by these techniques. Likewise, Spiral Display (SD), Small Multiples (SM), and Space-Time Cube (STC) all show relatively low similarity with techniques like Parallel Coordinates (PC) and Incidence Matrix (IM). The latter rely on line-based marks and structured layouts that emphasize alignment of visual regions, which may explain the difference. Such patterns suggest that visual encoding and layout can influence similarity scores as strongly as the categorization with respect to data facets.

Next, we assess how well the model-driven similarity scores align with human expert intuition to better understand the shaping of similarities.

\subsection{Similarity According to Experts}
\label{subsec:exp-results}

Based on the three expert similarity scores per pair of visualization techniques, we computed for each pair the average similarity score $[1.0-5.0]$ and the variance $[0.0-3.0]$. The average scores help us rank technique pairs in terms of perceived intuitive similarity, while the variances reveal where expert intuition diverged.

The expert-driven similarity scores are shown in the heatmap in \cref{fig:sim-model-expert} (right). Again, light yellow indicates low similarity and dark blue stands for high similarity. The heatmap reveals notable patterns similar to the ones in the model-driven heatmap. For example, Node-Link Diagram (NLD) and Scatter Plot (SP) are also rated as similar by the experts, not a strong though. At the same time, dissimilarities appear to be stronger (many light yellow cells with low scores), particularly for pairs involving techniques that visualize different data facets. For example, Colored Map (CM), Small Multiples (SM), Space-Time Cube (STC), and Network Map (NM) are rated as rather dissimilar to all other techniques, resulting in a pronounced cluster of techniques for spatial data S. Further expert responses suggest that their intuition was also influenced by marks and visual channels. For example, Scatter Plot (SP) and Spiral Display (SD), despite representing different data facets (attributes A vs. time T), still received a moderate similarity score of 3.7. Experts pointed to the use of orientation and area marks in Spiral Display (SD), along with the use of size to represent data in Scatter Plot (SP), as the main factors contributing to their perceived visual similarity. Likewise, Parallel Coordinates (PC) and Time Wheel (TW), which differ in axis orientation (Cartesian vs. radial), have a quite high similarity of 4.7, likely due to their common use of line marks.

The variance among expert similarity ratings is depicted in the heatmap in \cref{fig:var-expert}, where darker blue indicates higher variance, while lighter blue reflects lower variance in expert responses. In general, expert agreement was strong across most pairs, particularly among more basic techniques. A lower variance suggests a shared intuitive understanding of visual or structural similarity, while a higher variance indicates conflicting interpretations. Experts may have intuitively focused on different visual features, such as layout or mark shape, or interpreted the structural intent in different ways. For example, the pair of Node-Link Diagram (NLD) vs. Parallel Coordinates (PC) exhibited a relatively high variance of 3.0. Two experts rated them as dissimilar 1.0, while another gave a similarity score of 4.0. The dissimilarity was attributed to differences in shape, while the high similarity score was motivated by shared visual appearance. Such strong disagreements were relatively rare and tended to occur for pairs of techniques that either do or do not show the same data facets or use or do not use the same marks and visual channels. This suggests that data facets and visual appearance might be conflicting factors when judging visualization similarity. On the other hand, Parallel Coordinates (PC) vs. Scatter Plot (SP) had a low variance of 0.3, suggesting consistent perceptions of similarity.

Comparing model-driven and expert-driven heatmaps in \cref{fig:sim-model-expert}, discrepancies appeared in certain pairs where the model indicated similarity but expert ratings varied, such as in the case of Scatter Plot (SP) vs. Bar Table (BT). This highlights the need for different perspectives on similarity: The model perspective provides a systematic view based on technique properties, while the expert perspective reflects a more ad-hoc and intuitive contextual reasoning.

\subsection{Pathways of Similarity}

\begin{figure*}[t]
    \centering
    \includegraphics[width=\textwidth]{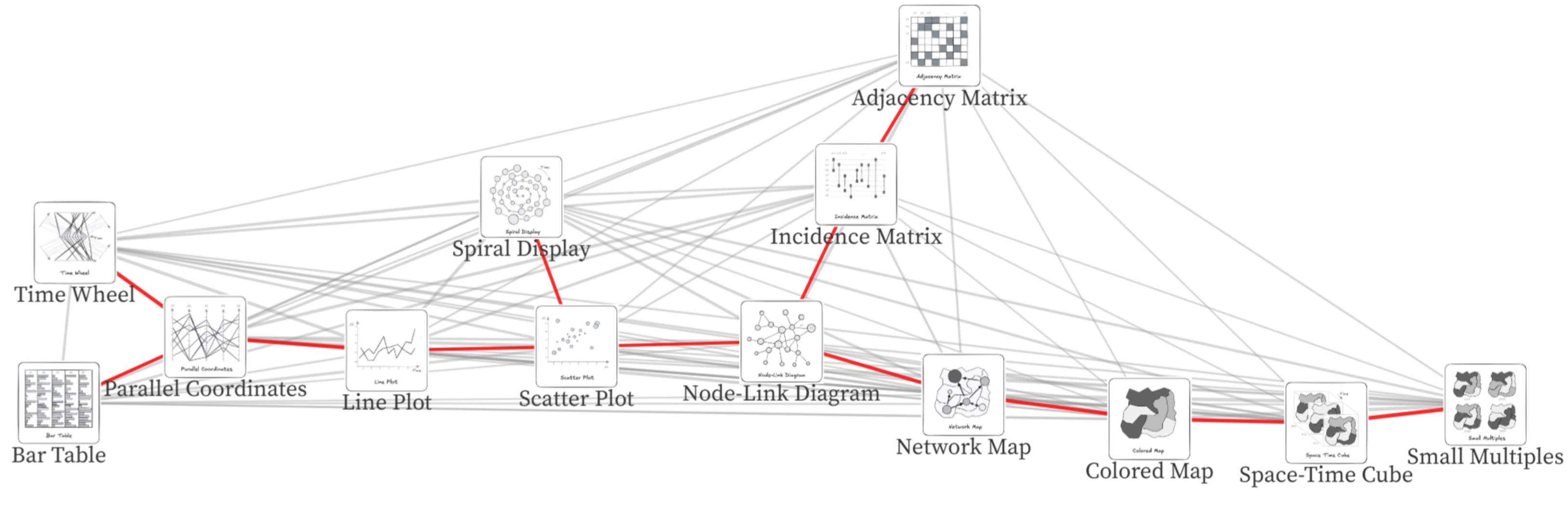}
    \caption{Minimum spanning tree illustrating the strongest expert-driven similarity between visualization techniques.}
    \label{fig:mst_expert}
\end{figure*}

Using the similarity scores from the previous sections, we can further explore the overall structure of relationships between visualization techniques. In particular, we were interested in how the visualization techniques are connected in terms of their similarity. We could ask, given a set of visualization techniques and their pairwise similarities, what is the minimum spanning tree (MST) over the similarity matrix? The MST will give us a structure in which edges connect the visualization techniques that are most similar, and any technique is reachable from any other technique via a high-similarity pathway through the MST. Relating back to the introduction, such an MST could provide insight into which techniques could or should be arranged next to each other in a multi-view visualization system or which techniques should be taught in which order to benefit from analogies and easy knowledge transfer.

We employed Kruskal's algorithm \cite{Kruskal56Spanning} to generate two MSTs. \cref{fig:mst_model} and \cref{fig:mst_expert} show the MSTs derived from the model-driven and the expert-driven similarity, respectively. In both figures, thicker red edges indicate the strongest similarity paths forming the backbone of the MST, while lighter gray edges represent the fully connected similarity graph of all possible paths.

The comparison between the expert-based and model-based MSTs reveals both shared pathways and distinct centralities among visualization techniques. Both trees consistently connect Colored Map through Space-Time Cube to Small Multiples, reflecting shared area-based encoding across different layouts. However, the two trees differ in which techniques emerge as central connectors. In the expert-driven MST, Node-Link Diagram serves as a key hub, linking multiple techniques through perceptual similarities and interpretive familiarity. In contrast, the model-driven MST positions Incidence Matrix more in the center, connecting area-based and line-based techniques through structural encodings. Additionally, Colored Map plays an important role in the model-driven MST, linking Spiral Display and Time Wheel, suggesting that its flexibility in encoding both spatial and time data is structurally significant.
\section{Conclusion}

With our work, we gained first insights into the similarity of visualization techniques. We employed two complementary methods that capture different aspects of similarity, including data-facets, visual and structural properties, as well as ad-hoc expert intuition. The obtained preliminary results already hint at some interesting relations between visualization techniques. Going beyond plain similarity values, we also extracted minimum spanning trees to define pathways through our collection of visualization techniques.

The biggest limitation of our work so far is its small scale in terms of number of techniques, model expressiveness, and number of experts. The natural next step would be to scale everything up. Therefore, future work could analyze a larger set of visualization techniques, expand the model expressiveness of our model-driven approach, maybe even considering grammars and SVG scene graphs, and conduct a larger-scale expert study.

Moreover, different notions of similarity should be investigated more systematically to build a more solid theoretical basis for understanding similarity of visualization techniques. Here, we considered mainly data facets and visual appearance of visualization techniques. Yet, also visualization tasks, interactivity, and visualization literacy play important roles in the perceived similarity of visualization techniques. Including these factors would be a formidable research challenge.

\acknowledgments{
This work was conducted in the scope of the project ``iV-Morph: Interactive Visual Metamorphosis for Multi-view Data Exploration'' funded by the Deutsche Forschungsgemeinschaft (DFG) -- German Research Foundation under grant number [\href{https://gepris.dfg.de/gepris/projekt/514630063?language=en}{514630063}].}

\bibliographystyle{abbrv-doi-hyperref}

\bibliography{main}

\begin{thebibliography}{10}

\bibitem{Aigner23TimeViz2nd}
\href{https://doi.org/10.1007/978-1-4471-7527-8}{W.~Aigner, S.~Miksch,
  H.~Schumann, and C.~Tominski}.
\newblock \href{https://doi.org/10.1007/978-1-4471-7527-8}{{\em {Visualization
  of Time-Oriented Data}}}.
\newblock \href{https://doi.org/10.1007/978-1-4471-7527-8}{Springer},
  \href{https://doi.org/10.1007/978-1-4471-7527-8}{second edition ed.},
  \href{https://doi.org/10.1007/978-1-4471-7527-8}{2023}.
  \href{https://doi.org/10.1007/978-1-4471-7527-8}
{doi: {{%
10\hspace{.1pt}\discretionary{.}{%
}{.}\hspace{.4pt}1007\discretionary{/}{%
}{/}978\discretionary{%
}{-}{-}1\discretionary{%
}{-}{-}4471\discretionary{%
}{-}{-}7527\discretionary{%
}{-}{-}8}}}


\bibitem{Beck14DynGraphVis}
\href{https://doi.org//10.2312/eurovisstar.20141174}{F.~Beck, M.~Burch,
  S.~Diehl, and D.~Weiskopf}.
\newblock \href{https://doi.org//10.2312/eurovisstar.20141174}{{The State of
  the Art in Visualizing Dynamic Graphs}}.
\newblock \href{https://doi.org//10.2312/eurovisstar.20141174}{In R.~Borgo,
  R.~Maciejewski, and I.~Viola, eds., {\em EuroVis - STARs}}.
  \href{https://doi.org//10.2312/eurovisstar.20141174}{The Eurographics
  Association}, \href{https://doi.org//10.2312/eurovisstar.20141174}{2014}.
  \href{https://doi.org//10.2312/eurovisstar.20141174}
{doi: {{%
\discretionary{/}{%
}{/}10\hspace{.1pt}\discretionary{.}{%
}{.}\hspace{.4pt}2312\discretionary{/}{%
}{/}eurovisstar\hspace{.1pt}\discretionary{.}{%
}{.}\hspace{.4pt}20141174}}}


\bibitem{Chen24Typology}
\href{https://doi.org/10.48550/ARXIV.2403.05594}{J.~Chen, P.~Isenberg, R.~S.
  Laramee, T.~Isenberg, M.~Sedlmair, T.~M{\"{o}}ller, and R.~Li}.
\newblock \href{https://doi.org/10.48550/ARXIV.2403.05594}{An image-based
  typology for visualization}.
\newblock \href{https://doi.org/10.48550/ARXIV.2403.05594}{{\em CoRR}},
  \href{https://doi.org/10.48550/ARXIV.2403.05594}{abs/2403.05594},
  \href{https://doi.org/10.48550/ARXIV.2403.05594}{2024}.
  \href{https://doi.org/10.48550/ARXIV.2403.05594}
{doi: {{%
10\hspace{.1pt}\discretionary{.}{%
}{.}\hspace{.4pt}48550\discretionary{/}{%
}{/}ARXIV\hspace{.1pt}\discretionary{.}{%
}{.}\hspace{.4pt}2403\hspace{.1pt}\discretionary{.}{%
}{.}\hspace{.4pt}05594}}}


\bibitem{Chen21MultiView}
\href{https://doi.org/10.1109/TVCG.2020.3030338}{X.~Chen, W.~Zeng, Y.~Lin,
  H.~M. Al{-}Maneea, J.~Roberts, and R.~Chang}.
\newblock \href{https://doi.org/10.1109/TVCG.2020.3030338}{Composition and
  configuration patterns in multiple-view visualizations}.
\newblock \href{https://doi.org/10.1109/TVCG.2020.3030338}{{\em IEEE
  Transactions on Visualization and Computer Graphics}},
  \href{https://doi.org/10.1109/TVCG.2020.3030338}{27(2):1514--1524},
  \href{https://doi.org/10.1109/TVCG.2020.3030338}{2021}.
  \href{https://doi.org/10.1109/TVCG.2020.3030338}
{doi: {{%
10\hspace{.1pt}\discretionary{.}{%
}{.}\hspace{.4pt}1109\discretionary{/}{%
}{/}TVCG\hspace{.1pt}\discretionary{.}{%
}{.}\hspace{.4pt}2020\hspace{.1pt}\discretionary{.}{%
}{.}\hspace{.4pt}3030338}}}


\bibitem{Hadlak15multifaceted}
\href{https://doi.org/10.2312/eurovisstar.20151109}{S.~Hadlak, H.~Schumann, and
  H.-J. Schulz}.
\newblock \href{https://doi.org/10.2312/eurovisstar.20151109}{{A Survey of
  Multi-faceted Graph Visualization}}.
\newblock \href{https://doi.org/10.2312/eurovisstar.20151109}{In R.~Borgo,
  F.~Ganovelli, and I.~Viola, eds., {\em Proceedings of the Eurographics / IEEE
  Conference on Visualization (EuroVis) - STAR}}.
  \href{https://doi.org/10.2312/eurovisstar.20151109}{Eurographics
  Association}, \href{https://doi.org/10.2312/eurovisstar.20151109}{2015}.
  \href{https://doi.org/10.2312/eurovisstar.20151109}
{doi: {{%
10\hspace{.1pt}\discretionary{.}{%
}{.}\hspace{.4pt}2312\discretionary{/}{%
}{/}eurovisstar\hspace{.1pt}\discretionary{.}{%
}{.}\hspace{.4pt}20151109}}}


\bibitem{Kruskal56Spanning}
\href{https://doi.org/10.2307/2033241}{J.~B. Kruskal}.
\newblock \href{https://doi.org/10.2307/2033241}{{On the Shortest Spanning
  Subtree of a Graph and the Traveling Salesman Problem}}.
\newblock \href{https://doi.org/10.2307/2033241}{{\em Proceedings of the
  American Mathematical Society}},
  \href{https://doi.org/10.2307/2033241}{7(1):48--50},
  \href{https://doi.org/10.2307/2033241}{1956}.
  \href{https://doi.org/10.2307/2033241}
{doi: {{%
10\hspace{.1pt}\discretionary{.}{%
}{.}\hspace{.4pt}2307\discretionary{/}{%
}{/}2033241}}}


\bibitem{Munzner14VAD}
\href{https://doi.org/10.1201/b17511}{T.~Munzner}.
\newblock \href{https://doi.org/10.1201/b17511}{{\em Visualization Analysis and
  Design}}.
\newblock \href{https://doi.org/10.1201/b17511}{{A.K.} Peters visualization
  series}. \href{https://doi.org/10.1201/b17511}{A {K} Peters},
  \href{https://doi.org/10.1201/b17511}{2014}.
  \href{https://doi.org/10.1201/b17511}
{doi: {{%
10\hspace{.1pt}\discretionary{.}{%
}{.}\hspace{.4pt}1201\discretionary{/}{%
}{/}b17511}}}


\bibitem{Ruchikachorn15Learning}
\href{https://doi.org/10.1109/TVCG.2015.2413786}{P.~Ruchikachorn and
  K.~Mueller}.
\newblock \href{https://doi.org/10.1109/TVCG.2015.2413786}{{Learning
  Visualizations by Analogy: Promoting Visual Literacy through Visualization
  Morphing}}.
\newblock \href{https://doi.org/10.1109/TVCG.2015.2413786}{{\em IEEE
  Transactions on Visualization and Computer Graphics}},
  \href{https://doi.org/10.1109/TVCG.2015.2413786}{21(9):1028--1044},
  \href{https://doi.org/10.1109/TVCG.2015.2413786}{2015}.
  \href{https://doi.org/10.1109/TVCG.2015.2413786}
{doi: {{%
10\hspace{.1pt}\discretionary{.}{%
}{.}\hspace{.4pt}1109\discretionary{/}{%
}{/}TVCG\hspace{.1pt}\discretionary{.}{%
}{.}\hspace{.4pt}2015\hspace{.1pt}\discretionary{.}{%
}{.}\hspace{.4pt}2413786}}}


\bibitem{Schulz11Tree}
\href{https://doi.org/10.1109/MCG.2011.103}{H.-J. Schulz}.
\newblock \href{https://doi.org/10.1109/MCG.2011.103}{Treevis.net: A tree
  visualization reference}.
\newblock \href{https://doi.org/10.1109/MCG.2011.103}{{\em IEEE Computer
  Graphics and Applications}},
  \href{https://doi.org/10.1109/MCG.2011.103}{31(6):11--15},
  \href{https://doi.org/10.1109/MCG.2011.103}{2011}.
  \href{https://doi.org/10.1109/MCG.2011.103}
{doi: {{%
10\hspace{.1pt}\discretionary{.}{%
}{.}\hspace{.4pt}1109\discretionary{/}{%
}{/}MCG\hspace{.1pt}\discretionary{.}{%
}{.}\hspace{.4pt}2011\hspace{.1pt}\discretionary{.}{%
}{.}\hspace{.4pt}103}}}


\bibitem{Schulz15Preset}
\href{https://doi.org/https://doi.org/10.1016/j.jvlc.2015.09.004}{H.-J. Schulz
  and S.~Hadlak}.
\newblock
  \href{https://doi.org/https://doi.org/10.1016/j.jvlc.2015.09.004}{Preset-based
  generation and exploration of visualization designs}.
\newblock
  \href{https://doi.org/https://doi.org/10.1016/j.jvlc.2015.09.004}{{\em
  Journal of Visual Languages \& Computing}},
  \href{https://doi.org/https://doi.org/10.1016/j.jvlc.2015.09.004}{31:9--29},
  \href{https://doi.org/https://doi.org/10.1016/j.jvlc.2015.09.004}{2015}.
  \href{https://doi.org/10.1016/j.jvlc.2015.09.004}
{doi: {{%
10\hspace{.1pt}\discretionary{.}{%
}{.}\hspace{.4pt}1016\discretionary{/}{%
}{/}j\hspace{.1pt}\discretionary{.}{%
}{.}\hspace{.4pt}jvlc\hspace{.1pt}\discretionary{.}{%
}{.}\hspace{.4pt}2015\hspace{.1pt}\discretionary{.}{%
}{.}\hspace{.4pt}09\hspace{.1pt}\discretionary{.}{%
}{.}\hspace{.4pt}004}}}


\bibitem{Tominski23TimeVizBrowser}
\href{https://browser.timeviz.net}{C.~Tominski and W.~Aigner}.
\newblock \href{https://browser.timeviz.net}{{The TimeViz Browser -- A Visual
  Survey of Visualization Techniques for Time-Oriented Data}}.
\newblock \href{https://browser.timeviz.net}{https://browser.timeviz.net},
  \href{https://browser.timeviz.net}{2023}.
\newblock \href{https://browser.timeviz.net}{Version 2.0}.

\bibitem{Tominski20IVDA}
\href{https://doi.org/10.1201/9781315152707}{C.~Tominski and H.~Schumann}.
\newblock \href{https://doi.org/10.1201/9781315152707}{{\em {Interactive Visual
  Data Analysis}}}.
\newblock \href{https://doi.org/10.1201/9781315152707}{AK Peters Visualization
  Series}. \href{https://doi.org/10.1201/9781315152707}{CRC Press},
  \href{https://doi.org/10.1201/9781315152707}{2020}.
  \href{https://doi.org/10.1201/9781315152707}
{doi: {{%
10\hspace{.1pt}\discretionary{.}{%
}{.}\hspace{.4pt}1201\discretionary{/}{%
}{/}9781315152707}}}


\bibitem{Winkler90String}
\href{https://eric.ed.gov/?id=ED325505}{W.~E. Winkler}.
\newblock \href{https://eric.ed.gov/?id=ED325505}{String comparator metrics and
  enhanced decision rules in the fellegi-sunter model of record linkage}.
\newblock \href{https://eric.ed.gov/?id=ED325505}{In {\em Proceedings of the
  Section on Survey Research Methods}},
  \href{https://eric.ed.gov/?id=ED325505}{pp. 354--359}.
  \href{https://eric.ed.gov/?id=ED325505}{American Statistical Association},
  \href{https://eric.ed.gov/?id=ED325505}{1990}.

\end{thebibliography}
\end{document}